 \documentclass[final,5p,times,twocolumn,authoryear]{elsarticle}

\usepackage{amssymb}
\usepackage{lipsum}
\usepackage{caption}
\usepackage{hyperref}
\journal{High Energy Astrophysics}

\begin{document}

\begin{frontmatter}



\title{Log-normal behavior of Optical Polarization in a sample of Blazars: Evidence for Multiplicative Magneto-hydrodynamic Processes in Relativistic Jets}


\author[1,2]{A. Singh\corref{cor1}}
\ead{sarchanaa019@gmail.com}
\author[2]{A. Tolamatti}
\author[2,1]{K. K. Singh\corref{cor2}}
\ead{kksastro@barc.gov.in}
\author[3]{A. C. Gupta}
\author[2]{B. Ghosal}
\author[2,1]{K. K. Yadav}

\cortext[cor1]{Corresponding author}
\cortext[cor2]{Co-corresponding author}

\address[1]{Homi Bhabha National Institute, Anushakti Nagar, Mumbai, 400094, India}

\address[2]{Astrophysical Sciences Division, Bhabha Atomic Research Centre, Trombay, 400085, India}

\address[3]{Aryabhatta Research Institute of Observational Sciences (ARIES), Manora Peak, Nainital 263001, India}

\begin{abstract}

Optical synchrotron polarization is one of the most direct probes of the magnetic-field structure and particle-acceleration processes in the relativistic jets of blazars. Although long-term optical flux variability is often attributed to multiplicative stochastic processes because of its log-normal behavior, but the physical origin of optical polarization variability remains less understood. In this study, we investigate the long-term optical flux and polarization variability of a sample of blazars using nearly a decade of photo-polarimetric observations from the Steward Observatory. We find that, like the optical flux, the optical degree of polarization is generally well described by a log-normal distribution and shows an approximately linear relationship between its variability amplitude and mean value. We interpret this behavior within a magneto-hydrodynamic (MHD) framework in which the magnetic-field undergoes repeated stochastic changes driven by shocks, turbulence, velocity shear, and large-scale jet dynamics. The cumulative effect of these processes can naturally generate multiplicative variability, leaving similar statistical signatures in both the synchrotron flux and the degree of polarization. Our results provide new observational evidence that the long-term statistical behavior of optical flux and polarization may arise from a common set of MHD processes associated with magnetic-field evolution. 
% Optical polarization therefore offers an additional means of constraining the physical mechanisms responsible for variability, magnetic-field restructuring, and particle acceleration in relativistic jets.

\end{abstract}




\end{frontmatter}

%\tableofcontents



\section{Introduction}

Blazars constitute the most extreme subclass of radio-loud active galactic nuclei (AGNs), where one of the two relativistic jets pointing towards the earth, i.e., with a viewing angle of less than $10^{\circ}$ \citep{Urry_1995, Padovani_2017, Saikia_2022}. These extragalactic objects are mainly characterized by the Doppler-boosted non-thermal emission ranging from radio to TeV energies originating from a relativistic jet. The observed properties of the broadband non-thermal emission depend on the jet geometry and relativistic beaming effects. Traditionally, blazars are categorized into two subgroups depending on equivalent width of the optical emission lines: BL Lacertae (BL Lacs) objects with weak or no emission lines and Flat Spectrum Radio Quasars (FSRQs) with broad emission lines. One of the defined features of the blazars is the high level of variability in flux and polarization on various timescales, from minutes to years in all wavebands. Despite many decades of observations, the physical origin of this underlying mechanism remains less understood. 

Optical polarimetry programs spanning long timescales have shown that blazars display a wide range of behavior in their polarization properties, from stable ones to stochastic changes along with considerable rotation of the electric vector polarization angle (EVPA). Prolonged monitoring observations, specifically those performed by the Steward Observatory and the RoboPol project, have shown that when blazars are experiencing a phase of strong optical activity, this is usually accompanied by pronounced changes in the polarization degree (PD) and EVPA; however, these connections are not always present \citep{Abdo_2010, Larionov_2013, Gaur_2014}. Flares with an increase in PD suggest domination of an ordered magnetic-field during this period of time, while in the case of flares with small or no changes in polarization, this indicates the presence of several physical processes contributing to the variation \citep{Smith_2009, Blinov_2015, Blinov_2016, Pavlidou_2014}. Recent analyses based on the Steward Observatory spectro-polarimetric monitoring program have provided detailed investigations of long-term optical flux and polarization variability in individual blazars, demonstrating the importance of continuous polarimetric observations for understanding jet physics \citep{Rajput_2022, Pandey_2022, Singh_2025, Archana_2026}. More recent optical polarization studies further indicate that both ordered and turbulent magnetic-field components contribute to the observed variability, with turbulence becoming increasingly important during active states \citep{Alok_2017, Alok_2023}. Several models reported in the literature suggest that optical flux and polarization variability in blazars can arise from both intrinsic and extrinsic mechanisms associated with the source. In the intrinsic scenario, variability is primarily attributed to physical processes occurring within the central engine, particularly fluctuations in the relativistic jets and the accretion disk surrounding the supermassive black hole. This include interactions between moving shocks and turbulent regions, fluctuations in the magnetic-field, and variations in particle acceleration mechanisms \citep{Wagner_Witzel_1995, Bhatta_2013, Marscher_2014, Calafut_2015, Bitan_2024}. Such processes naturally introduce stochastic variability in the emitted radiation. In contrast, extrinsic processes are related to geometrical effects of the emission region rather than changes in the intrinsic physical scenario. These processes include variations in Doppler boosting due to small changes in the orientation of the emission blob with respect to the observer's line of sight and bending of jet \citep{Camenzind_1992, Raiteri_2017, Larionov_2010, Raiteri_2013}. 

A number of models have been developed in order to describe flux and polarization variations in blazars. In the transverse shock model, the propagation of relativistic shocks causes compression of the magnetic-field and leads to the enhancement of the synchrotron radiation and higher polarization \citep{Marscher_Gear_1985, Hughes_1985}. On the other hand, the existence of large scale helical magnetic-field can easily account for smooth EVPA rotations and systematic polarization changes due to geometrical effects and viewing angle variations \citep{Lyutikov_2005, Marscher_2008, Zhang_2014, Zhang_2015}. Turbulence models assume that the random variations of the flux and polarization are related to the turbulence plasma cells passing the standing shocks in the jet \citep{Marscher_2014}. Finally, the recent studies have shown that the magnetic reconnection can be considered another promising process responsible for fast particle acceleration, dissipation of local energy and variability on multiple wavelengths. It should be noted that the co-existence of all these phenomena implies that the variability must be driven by both ordered and turbulent parts of the magnetic-field. Statistical properties of light curves such as flux distribution, rms-flux relation, and properties of power spectra serve as important diagnostics of the physics of relativistic jets. If the observed variability arises from the superposition of many independent and additive processes, then in such circumstances, the Central Limit Theorem (CLT) predicts that the resulting flux distribution should follow a Gaussian profile, whose width tells about the degree of variability. Although various studies have investigated flux distributions in blazar light curves and the result is quite interesting, i.e., the flux distribution deviates significantly from Gaussian behavior, instead follow log-normality. This behavior has been observed in many blazars in the optical, X-ray, and $\gamma$-ray bands \citep{Uttley_2005, Giebels_2009, Sinha_2016, Zahir_2018, Bhatta_2020, Bhatta_2021}. In a log-normal distribution, the logarithm of the flux is normally distributed, implying that the variability arises from multiplicative processes rather than simple additive fluctuations. The log-normality of flux distributions is generally associated with non-linear variability processes and is often followed by a linear rms-flux relation, implying the variations follow a log-normal behavior. 

While there have been many studies about the statistical properties of flux variability, relatively little effort has gone into determining if the level of polarization is also distributed according to the log-normal form. Since the optical polarization is due to synchrotron radiation and directly dependent upon magnetic-field geometry and particle acceleration within the jets, any physical multiplicative process that influences flux variability may influence the distribution of the level of polarization as well. Therefore, the study of the statistical properties of both flux and polarization allows us to explore the physics of jets even further. Inspired by such a possibility, we conduct a thorough statistical analysis of optical photometry and polarization measurements of a sample of BL Lac objects and FSRQs for almost a decade obtained with the Spectro-Polarimeter (SPOL) at Steward Observatory. The findings put new constraints on the origin of optical variability and on the effect of multiplicative stochastic processes within jets. The remainder of this paper is organized as follows. Section \ref{sec:2} presents the observational data set, Section \ref{sec:3} describes the statistical framework of the log-normal distribution, Section \ref{sec:4} introduces the polarization models adopted in this study, Section \ref{sec:5} reports the results, and finally Section \ref{sec:6} and Section \ref{sec:7} discusses and summarizes the conclusions of this work, respectively. 

\section{Data set}
\label{sec:2}

In this work, we carry out a comprehensive statistical investigation of the long-term optical V-band flux along with the distribution of the PD for a sample of 10 blazars (3 BL Lacs and 7 FSRQs). The SPOL at the Steward Observatory of the University of Arizona provides measurements of the optical linear polarization of the sources in the wavelength
range of $\lambda$=500-700 nm \citep{Smith_2009}\footnote{\url{http://james.as.arizona.edu/~psmith/Fermi/DATA/individual.html}}. These data have been publicly released under the \textit{Multiwavelength Observing Support Programs} to support the \textit{Fermi science objectives}\footnote{\url{https://fermi.gsfc.nasa.gov/ssc/observations/multi/programs.html}}. The data sets used in this work span approximately a decade, from 2008 to 2018, allowing us to probe the long-term variability of flux and polarization. We restricted our sample to sources with more than 100 independent measurements of flux and polarization. This selection criterion helps to avoid small statistical biases. 
In addition to the flux and PD, the SPOL data set also provides the normalized Stokes parameters, q = Q/I and u = U/I, from which the PD is calculated as PD = $\sqrt{(q^2 \ + \ u^2)}$. Because PD is a positive-definite quantity obtained from the quadrature sum of q and u, it can be affected by a well-known positive bias, particularly at low signal-to-noise ratio, which can itself produce a non-Gaussian, right-skewed distribution even when the underlying polarization is statistically constant \citep{Rice_1944}. To test whether the log-normal behavior of PD reported in this work reflects genuine multiplicative physical processes rather than this measurement bias, we additionally analyze the distributions of the absolute normalized Stokes parameters, $|q|$ and $|u|$, for the same sample of blazars. 

\section{Log-normal distribution}
\label{sec:3}

The variability observed in blazars over a wide range of timescales and amplitudes is highly complex. Several studies in the literature suggest that such variability may arise from a cascading effect of different stochastic processes operating within the relativistic jet and the accretion flow. This leads to multiplicative fluctuations in the observed emission. As a result, the statistical properties of blazar light curves are expected to follow log-normal distribution instead of Gaussian behavior. \\

Let f(x) represents the product of the flux contributions originating from a large number of independent isotropic regions ($ \rm x_i$), then

\begin{center}
    f(x) = $ \rm \prod_{i=1}^{i=n} x_i $\\
    log\{f(x)\} = log$ \rm \{\prod_{i=1}^{i=n} x_i\}$  \\
    log\{f(x)\} = $ \rm \sum_{i=1}^{i=n} $ log $ \rm x_i$
\end{center}

So, log\{f(x)\} is the summation of log$\rm (x_i)$. According to the CLT, the logarithmic value of flux distribution f(x) follows a normal distribution, this in turn implies that f(x) follows a distribution known as, \textit{log-normal distribution}. Hence, the multiplicative processes naturally generate log-normality, with a characteristics tailed and right-skewed distribution. The probability distribution function (PDF) of the log-normal distribution is defined as,

\begin{equation}
   \rm f_{log-normal}(x) = \frac{1}{x\sigma \sqrt{2\pi}} exp \left[ - \frac{(\ln x - \mu)^2}{2\sigma^2} \right]
\end{equation}

where $\mu$ and $\sigma$ are the mean value and standard deviation of $\ln \rm x$, respectively. These parameters denote the location parameter and scale parameter of the distribution in logarithmic space. 

\section{Blazar synchrotron polarization models}
\label{sec:4}

The observed variations in PD and EVPA can reveal important information about the geometry, configuration, and temporal evolution of the magnetic-field embedded within the emitting region. Over the past years, several theoretical models have been proposed to explain the wide-range of such long-term polarization behavior observed in blazars. In this section, we briefly describe three of the most widely used polarization models:

\subsection{Transverse shock model} 

The transverse shock model is one of the most widely used frameworks for describing flux and polarization variability in blazars \citep{Marscher_Gear_1985, Hughes_1985}. In this model, a relativistic shock propagates either perpendicular or nearly perpendicular to the jet axis and compresses the initially turbulent magnetic-field in the synchrotron emitting plasma. This compression leads to a partial alignment of the magnetic-field lines within the plane of the shock \citep{Laing_1980, Hughes_1985}. As the ordering of the magnetic-field grows, particle acceleration and magnetic-field amplification enhance the synchrotron emission. This results in a simultaneous increase in both the observed flux and PD \citep{Marscher_Gear_1985}. Because the compressed and ordered magnetic-field has a certain preferred orientation, the EVPA is expected to remain slightly stable or showing a small changes during the flare \citep{Hughes_1985, Marscher_2008}. As the shocked region expands and evolves with time, the ordered magnetic-field gradually returns to its original turbulent state. Consequently, both the flux and PD decrease to their initial values. 

\subsection{Helical magnetic-field model}

Another widely used framework for interpreting the variations in the blazar flux and polarization is the helical magnetic-field model. In this framework, the relativistic jet is threaded by a large-scale helical magnetic-field that is established via the processes of jet launching and collimation \citep{Blandford_1983, Blandford_1977, Lyutikov_2005, Marscher_2014}. The observed synchrotron emission is produced by the relativistic electrons spiraling along this large-scale magnetic-field, while PD and EVPA are determined by the projection of the magnetic-field on the plane of the sky \citep{Lyutikov_2005, Marscher_2008}. Variability can naturally arise when various processes, such as shocks, magnetic-reconnection events, or turbulence within the plasma flow propagate through the helical field. These processes may alter the relative strength of the magnetic-field components (toroidal and poloidal), and can also change the Doppler factor through geometrical changes in the system \citep{Marscher_2008, Zhang_2014, Zhang_2015}. As emitting plasma follows a curved or spiral trajectory through the helical field, the projected magnetic-field direction changes gradually. This can produce smooth EVPA rotations accompanied by systematic variations in the PD and synchrotron flux, without requiring the behavior to be dominated by purely stochastic fluctuations \citep{Marscher_2008, Abdo_2010, Zhang_2015}. The model can reproduce several well-known characteristics of the polarization pattern of blazars, including long-duration EVPA rotations, polarization swings associated with the multi-wavelength flares, and enhanced PD during high flux states \citep{Marscher_2008, Abdo_2010, Blinov_2015, Blinov_2016}. Therefore, this model provides a plausible physical framework for connecting jet geometry, magnetic-field structure, relativistic beaming, with the synchrotron emission.

\subsection{Turbulent Extreme Multi-Zone (TEMZ) model}

The stochastic, noise-like variability commonly observed in blazars indicates that turbulence may play an important role in the dynamics of relativistic jets \citep{Marscher_2014, Calafut_2015, Bhatta_2013}. To interpret this behavior, \citet{Marscher_2014} introduced the Turbulent Extreme Multi-Zone (TEMZ) model, in which the observed multi-wavelength flux and polarization variability arise from turbulent plasma propagating through the jet. In this model, the ambient jet plasma is divided into a large number of turbulent cells. Each cell contains relativistic electrons (and positrons), together with a locally uniform magnetic-field whose orientation varies randomly from one cell to another cell. As these cells traverse a standing shock, the particles are accelerated and produce radiation via synchrotron and inverse-Compton processes \citep{Marscher_2014}. The combined emission from various cells produces the observed variability. Furthermore, the co-existence of an ordered large-scale magnetic-field and a turbulent magnetic-field component can naturally account for both a preferred polarization orientation and the rapid fluctuations in flux, PD, and EVPA commonly observed in blazars \citep{Marscher_2014, Blinov_2015, Blinov_2016}. 

\section{Results}
\label{sec:5}

The decade-long temporal analysis of the polarimetric and photometric emissions from a sample of sources is shown in Figure \ref{fig:1}. A clear visual inspection indicates that these light curves are highly variable. In order to quantify this variability, we applied the Bayesian block algorithm to the one-day binned light curves  \citep{Scargle_2013}. The visual representation of this algorithm is shown in Figure \ref{fig:1}. For this analysis, a false alarm probability ($p_{0}$ = 0.01) is adopted, which ensures that the introduction to a new block is not due to statistical noise but reflects genuine variations of the optical flux in the V-band and PD\footnote{\url{https://docs.astropy.org/en/stable/api/astropy.stats.bayesian_blocks.html}}.

\subsection{\textbf{Variability analysis and correlations}}

We have carried out a detailed study aimed at characterizing the variability properties of the decade-long V-band optical flux and PD exhibited by the selected blazars. 

\subsubsection{Fractional variability amplitude}

\begin{figure}[ht!]
\centering
\includegraphics[width=9cm, height=17cm]{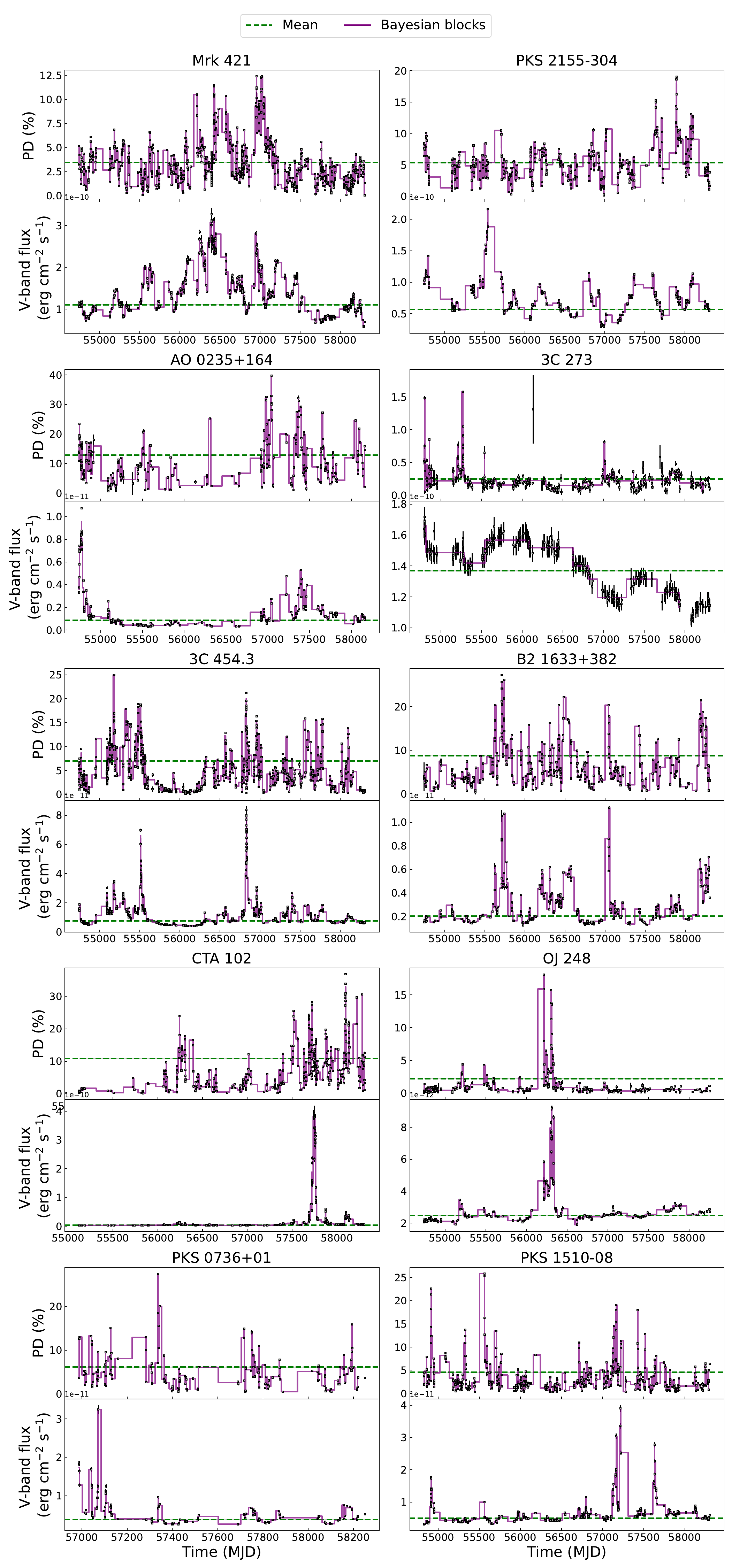}
\caption{The daily optical light curves of the blazar sample. The black squares represent the data points correspond to that panel.}
\label{fig:1}
\end{figure}

The long-term optical light curves of the V-band flux, along with PD in the wavelength range $\lambda$ = 500-700 nm, are shown in Figure \ref{fig:1}. Visually, it is evident that over long periods both the physical quantities are highly variable for all the sources. In order to investigate the statistical variability behavior of both quantities, we performed an intrinsic variability analysis using the fractional variability amplitude ($\rm F_{var}$), which measures the strength of variability by accounting for the measurement uncertainties in the observational data. This parameter provides a robust estimate of the intrinsic variability present in the flux and polarization time series. It is defined as \citep{Vaughan_2003, SINGH_2020},

\begin{equation}
\rm F_{var} = \sqrt{\rm \frac{S^2 - E^2}{F^2}}
\end{equation}

and the error in $\rm F_{\rm var}$ ($ \rm  \Delta F_{\rm var}$) is given by,

\begin{equation}
\rm {\Delta{F_{var}}} = \sqrt{\left( \sqrt{\frac{1}{ \rm 2N}}\left(\frac{ \rm E^2}{ \rm F^2 F_{var}} \right)^2 + \left( \frac{\rm E^2}{\rm N}\frac{1}{\rm F} \right)^2 \right)}  
\end{equation}

where $S^2$ is the variance, F is the mean of the data set, N is the number of measurements over a decade, and $E^2$ is the mean square measurement error, calculated by averaging the squares of the individual measurement uncertainties, as the following,

\[E^2=\frac{1}{N}\sum_{i=1}^{N}\sigma_{i}^{2}\]

where \(\sigma_{i}\) represents the measurement uncertainty for the \(i\)-th observation. The estimated values of $\rm F_{\rm var} \pm \Delta \rm F_{\rm var}$ for the optical flux in the V-band and PD of the selected blazars are represented in columns (4) and (5) of Table \ref{tab:1}, respectively. It is clearly evident that FSRQs are more variable than BL Lacs objects. 
In order to find the possible correlation between the PD and optical V-band flux, we perform a cross-correlation analysis using the z-transformed discrete correlation function (ZDCF) \citep{Alexander_1997}. The results of the ZDCF analysis for the sampled blazars are reported in Table \ref{tab:2}, and they exhibit positive correlation peaks of variable strength, with correlation coefficients ranging from approximately 0.53 to 0.93. The observed positive correlations may be associated with the changes in the magnetic-field configuration and orientation of the emission region within the relativistic jet. In particular, variations in the jet viewing angle can simultaneously influence the observed optical flux through Doppler boosting and the PD through changes in the projected magnetic-field geometry. Such behavior can be explained within the framework of helical magnetic-field and transverse shock models, as discussed by \citet{Archana_2026} and \citet{Raiteri_2013}. This signifies that the log-normality of the PD and optical flux may not comprise independent evidence of multiplicative variability, as both quantities can be influenced by common physical processes within the relativistic jets.

\begin{table*}[ht]
\centering
\caption{Derived values of the ($F_{\rm var}\pm\Delta F_{\rm var}$), F-test, RVA, and ANOVA-test for the V-band optical flux ($F_V$) and PD, together with the redshift ($z$) and class of the sample blazars.}
\label{tab:1}
\small
\begin{tabular}{lcc|cc|cc|cc|cc}
\hline
{Blazar name} &
{$z$} &
{Class} &
\multicolumn{2}{c}{$F_{\rm var}\pm\Delta F_{\rm var}$ (\%)} &
\multicolumn{2}{c}{F-test} &
\multicolumn{2}{c}{RVA} &
\multicolumn{2}{c}{ANOVA-test} \\
\cline{4-11}
&&&
$F_V$ & PD &
$F_V$ & PD &
$F_V$ & PD &
$F_V$ & PD \\
(1) & (2) & (3) & (4) & (5) & (6) & (7) & (8) & (9) & (10) & (11) \\
\hline
Mrk 421      & 0.031 & BL Lac & $40.41\pm0.09$  & $66.73\pm0.05$  & 330.31  & 1915.80 & 0.70 & 0.99 & 200.40 & 37.83 \\
PKS 2155$-$304 & 0.116 & BL Lac & $37.97\pm0.09$  & $55.46\pm0.06$  & 609.13  & 2075.01 & 0.77 & 0.98 & 17.80  & 37.44 \\
AO 0235+164 & 0.940 & BL Lac & $107.62\pm0.14$ & $69.85\pm0.27$  & 2527.95 & 243.99  & 0.95 & 0.97 & 12.32  & 15.66 \\
3C 273       & 0.158 & FSRQ   & $10.38\pm0.22$  & $71.44\pm1.04$  & 8.91    & 14.56   & 0.24 & 0.95 & 384.87 & 24.46 \\
3C 454.3     & 0.859 & FSRQ   & $89.02\pm0.15$  & $79.23\pm0.08$  & 586.96  & 1276.44 & 0.91 & 0.98 & 21.73  & 22.06 \\
B2 1633+382  & 1.810 & FSRQ   & $57.97\pm0.10$  & $74.89\pm0.16$  & 893.54  & 487.60  & 0.80 & 0.97 & 20.12  & 23.47 \\
CTA 102      & 1.037 & FSRQ   & $225.78\pm0.56$ & $82.85\pm0.08$  & 512.40  & 2500.76 & 0.99 & 0.99 & 16.00  & 49.99 \\
OJ 248       & 0.939 & FSRQ   & $41.70\pm0.15$  & $179.00\pm0.71$ & 308.60  & 241.92  & 0.66 & 0.99 & 23.31  & 19.91 \\
PKS 0736+01  & 0.189 & FSRQ   & $70.13\pm0.40$  & $81.02\pm0.28$  & 228.53  & 497.78  & 0.86 & 0.97 & 14.87  & 4.46 \\
PKS 1510$-$08 & 0.361 & FSRQ   & $63.39\pm0.17$  & $100.84\pm0.23$ & 365.10  & 402.47  & 0.86 & 0.98 & 16.06  & 6.83 \\
\hline
\end{tabular}
\end{table*}

\begin{table*}[ht]
\centering
\caption{ZDCF correlation coefficients and maximum-likelihood time lags for the blazar sample.}
\label{tab:2}
\setlength{\tabcolsep}{16.8pt}
\renewcommand{\arraystretch}{1.2} % Increase row spacing
\begin{tabular}{lll}
\hline
Blazar name & ZDCF & Peak likelihood \\
\hline
Mrk 421 & $0.680^{+0.145}_{-0.180}$  & 0.1969 \\

PKS 2155$-$304 & $0.890^{+0.045}_{-0.056}$ & 0.0931 \\

AO 0235+164 & $0.668^{+0.093}_{-0.106}$ & 0.0076  \\

3C 273 & $0.530^{+0.179}_{-0.208}$ & 0.0489 \\

3C 454.3 & $0.939^{+0.036}_{-0.049}$ & 0.2575 \\

B2 1633+382 &$0.739^{+0.057}_{-0.063}$ & 0.0575 \\

CTA 102 & $0.834^{+0.083}_{-0.109}$ & 0.0594 \\

OJ 248 & $0.719^{+0.032}_{-0.033}$ & 0.0343 \\

PKS 0736+01 & $0.835^{+0.091}_{-0.123}$ & 0.0384 \\

PKS 1510$-$08 & $0.756^{+0.068}_{-0.077}$ & 0.0817 \\

\hline

\end{tabular}
\end{table*}

\subsubsection{F-test}

The observed variability in both quantities is further examined using the F-test. The F-statistics compares the observed variance in the blazar light curve with the variance in the measurement uncertainties, i.e. \citep{Howell_1988},

\begin{equation}
  \rm  F = \frac{\sigma_{LC}^2}{\sigma^2}
\end{equation}

If the F-value exceeds a critical value of 6.636 for a chosen significance level corresponding to (N-1) degrees of freedom, it means the observed variations cannot be solely explained by observational errors and therefore represent intrinsic variability of the source \citep{Diego_2010}. The calculated values of the F-test for the V-band flux and PD are listed in columns (6) and (7) of Table \ref{tab:1}. It is clearly evident that all sources are highly variable throughout the sample. Furthermore, the generally larger F-values observed for PD indicate that polarization exhibits stronger intrinsic fluctuations than the optical flux, suggesting a highly dynamic evolution of the magnetic-field within the emission region.

\subsubsection{Relative variability amplitude}

The relative variability amplitude (RVA) is further calculated to quantify the amplitude of the observed variability, independent of the absolute brightness of the source, defined as \citep{Kovalev_2005, Singh2020, Singh2018},

\begin{equation}
  \rm  RVA = \frac{F_{max} - F_{min}}{F_{max} + F_{min}}
\end{equation}

where $\rm F_{max}$ and $\rm F_{min}$ denote the highest and lowest flux values. The RVA values ranges from 0 to 1, with larger values corresponding to stronger variability. The estimated RVA values for the V-band flux and PD are presented in columns (8) and (9) of Table \ref{tab:1}. The remarkably high RVA values of PD indicate that the polarization varies over nearly its entire period of observation.

\subsubsection{ANOVA-test}

To independently verify the presence of long-term variability, we also performed a one-way analysis of variance (ANOVA) test \citep{Diego_1998, Diego_2010}. This statistical method examines whether the mean values of different subsets of the light curve differ significantly from one another. The ANOVA statistics obtained for the optical flux and PD are listed in columns (10) and (11) of Table \ref{tab:1}. The calculated values are significantly larger than unity for all sources, demonstrating that both the optical flux and PD exhibit statistically significant long-term variations during the observational period. 

The agreement among these independent statistical techniques provides robust evidence that the optical emission and polarization properties of the sample blazars are highly variable over decade-long timescales. Such pronounced variability is naturally expected from stochastic processes operating within relativistic jets, including turbulence, shocks, magnetic-field evolution, and changes in Doppler boosting.

\subsection{Flux and polarization distribution}

\begin{figure}[ht!]
\centering
\includegraphics[width=0.45\textwidth]{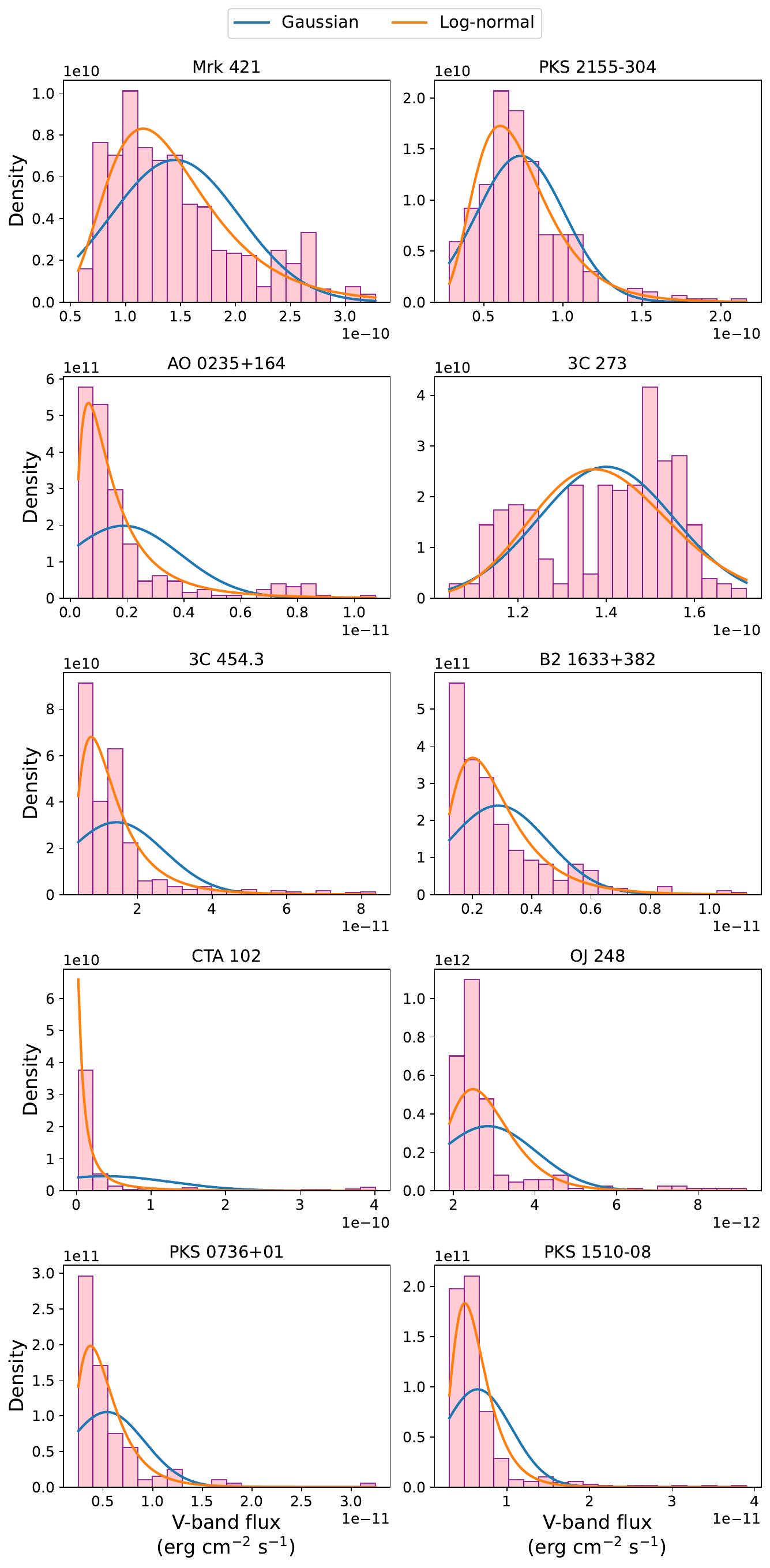}
\caption{V-band flux distribution of the sample of blazars. The blue and orange lines correspond to Gaussian and log-normal distribution, respectively.}
\label{fig:2}
\end{figure}

\begin{figure}[ht!]
\centering
\includegraphics[width=0.45\textwidth]{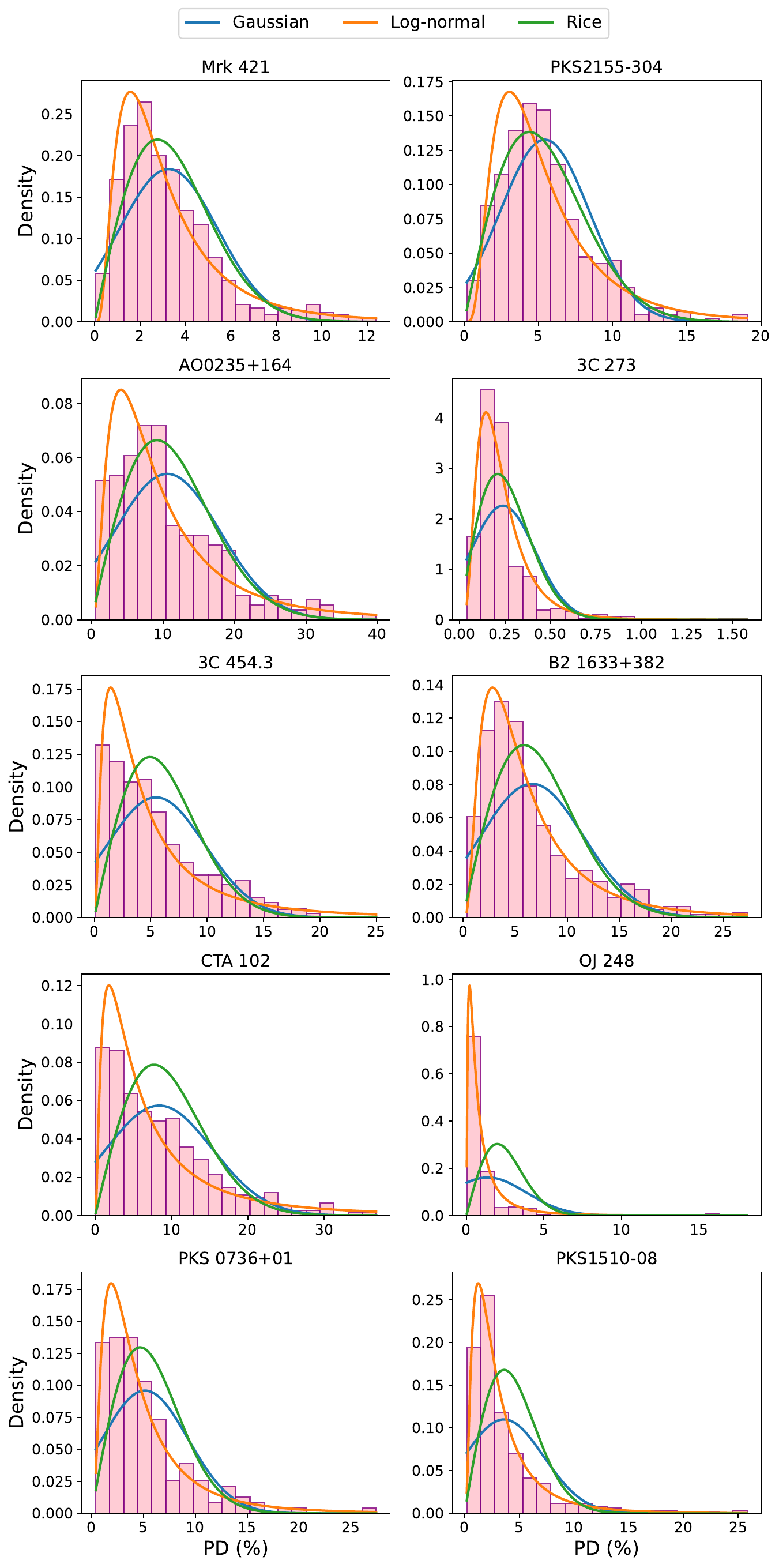}
\caption{Polarization distribution of the sample of blazars. The blue, orange and green lines correspond to Gaussian, log-normal and Rice distribution, respectively.}
\label{fig:3}
\end{figure}

\begin{figure}[ht!]
    \centering
    \includegraphics[width=1.0\linewidth]{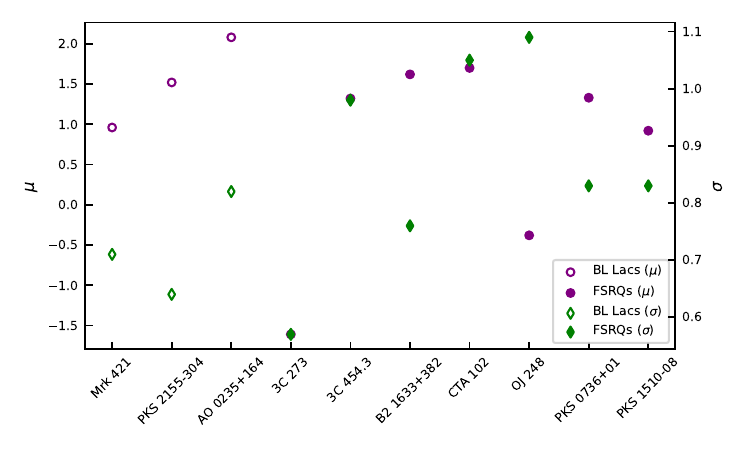}
    \caption{The best fit log-normal parameters, $\mu$ and $\sigma$, for the degree of linear polarization in a sample of blazars.}
    \label{fig:4}
\end{figure}

As discussed earlier, the variability in PD is intrinsically linked to the same physical mechanisms responsible for long-term flux variability. Accordingly, we examine the statistical distributions of both quantities. This joint investigation provides important insights into the origin and physical nature of the variability. It can probe the underlying stochastic mechanisms operating in relativistic jets, such as turbulence, magnetic-field fluctuations, and multiplicative or additive variability processes \citep{Uttley_2005}. Histograms of the V-band optical flux density and PD are constructed to determine the PDFs. The resulting distributions of the optical flux density and PD are modeled using Gaussian and log-normal distributions, as shown in Figures \ref{fig:2} and \ref{fig:3}, respectively, to determine which distribution best describes the observed variability. Since PD is a positive-definite quantity and is subject to a well-known positive bias, we additionally fit the Rice distribution to the observed PD distributions. The Rice distribution provides an appropriate statistical framework for describing the measured PD in the presence of measurement uncertainties \citep{Rice_1944}. The results show that the log-normal distribution provides a better description of the observed PD distribution than the Rice distribution. The best-fit log-normal parameters $\mu$ and $\sigma$, representing the mean and standard deviation of the natural logarithm of the observed PD, respectively, are presented for each source in Figure \ref{fig:4}.

\subsubsection{Akaike Information Criterion and Bayesian Information Criterion}

To statistically confirm which distribution better represents the observed data, we further calculated the \textit{Akaike Information Criterion (AIC)} and \textit{Bayesian Information Criterion (BIC)} for each considered distributions, which provide penalizing terms for large number of parameters and, therefore, provide a balance between goodness-of-fit and model complexity \citep{Akaike_1974}. These estimators are defined as,

\begin{equation}
\rm    AIC = 2k - 2 \ln (L)
\end{equation}

and,

\begin{equation}
\rm    BIC = k\ ln (n) - 2 \ln (L)
\end{equation}

where, k is the number of free parameters in the model, L represents the maximum likelihood, and n denotes the total number of data points. AIC is used to evaluate the relative quantity of a statistical model by estimating the error in the model fitting to the observed data, while BIC provides a criterion that help to identify the model that most appropriately describes the data. Statistically, the model with smaller AIC and BIC values is preferred. For flux distribution, the AIC and BIC values are reported in columns (2) and (3), and (6) and (7) of Table \ref{tab:2} for normal and log-normal distributions, respectively. And for PD, the AIC and BIC values are listed in columns (2) and (3), (7) and (8), and (12) and (13) of Table \ref{tab:3} for normal, log-normal and Rice distributions, respectively. It can be seen that the highlighted AIC and BIC values for the log-normal distribution turn out to be smaller than the other distributions for both the optical flux and PD. Therefore, the asymmetric log-normal curves, with a heavy tail extending towards the higher flux/polarization values, better represent the distribution of the V-band flux and PD.

\subsubsection{Anderson-Darling test}

In addition to the above statistical methods, we used the \textit{Anderson-Darling (AD) test}, which provides a measure of the agreement between the flux and polarization data with the fitted statistical models, and is particularly sensitive to the high-tails of the distribution \citep{Anderson_1952}. 
% This test evaluates whether the observed data are consistent with a given theoretical model.
The smaller the AD statistic, the better the consistency between the observed and modeled distributions. These statistical frameworks provide whether the optical flux and PD follow a Gaussian distribution, typically associated with linear stochastic processes, or a log-normal distribution, which corresponds to multiplicative or non-linear variability mechanisms operating in relativistic jets. The estimated values of the AD statistics for the V-band flux for each considered distributions are represented in columns (4) and (8) of Table \ref{tab:2} for Gaussian and log-normal distributions, and  for the degree of optical linear polarization are represented in columns (4), (9) and (14) of Table \ref{tab:3} for Gaussian, log-normal and Rice distributions, respectively. From the highlighted values of the AD test, it can be seen that the log-normal distribution is better consistent with the observed data than the Gaussian and Rice distributions. 

\subsubsection{Reduced chi-squared}

Furthermore, the reduced chi-squared ($\rm \chi_{r}^{2}$) is also calculated to evaluate goodness-of-fit test of the Gaussian, log-normal and Rice models. The corresponding values for the optical flux are listed in columns (5) and (9) of Table \ref{tab:2}, and for PD, along with the associated degrees of freedom (DoF) are presented in columns (5), (10) and (15) of Table \ref{tab:3}, respectively. The log-normal model systematically yields smaller $\rm \chi_r^2$ values than the Gaussian and Rice model for both the optical flux and PD. The only exceptions are the optical flux distribution of 3C 273 and PD distribution of PKS 2155-304, for which the Gaussian and Rice models provide a better agreement, respectively. Overall, the $\rm \chi_{r}^{2}$ analysis independently confirms the conclusions drawn from the AIC, BIC, and AD tests.

\begin{table*}[ht]
\centering
\caption{Comparison of Gaussian and log-normal fits to the V-band flux distribution of the blazar sample. The lower values of AIC, BIC, AD test, and $\chi_r^2$ corresponding to the preferred fit are highlighted in bold.}
\label{tab:3}
\setlength{\tabcolsep}{7pt}
\begin{tabular}{l|cccc|cccc}
\hline
{Blazar name} &
\multicolumn{4}{c}{Gaussian} &
\multicolumn{4}{c}{Log-normal} \\
\cline{2-5}\cline{6-9}
& AIC & BIC & AD & $\chi_r^2$
& AIC & BIC & AD & $\chi_r^2$ \\
(1) & (2) & (3) & (4) & (5) & (6) & (7) & (8) & (9) \\
\hline

Mrk 421      & -26564.99 & -26556.19 & 16.3187 & 10.603 &
\textbf{-26714.79} & \textbf{-26705.99} & \textbf{3.5596} & \textbf{4.834} \\

PKS 2155$-$304 & -14825.82 & -14818.26 & 5.9216 & 5.888 &
\textbf{-14905.01} & \textbf{-14897.45} & \textbf{0.9935} & \textbf{4.437} \\

AO 0235+164  & -12497.95 & -12490.95 & 28.1358 & 17.379 &
\textbf{-12814.15} & \textbf{-12807.15} & \textbf{2.2956} & \textbf{2.676} \\

3C 273       & \textbf{-14411.60} & \textbf{-14404.15} & \textbf{6.6221} & \textbf{15.426} &
-14396.13 & -14388.67 & 8.2105 & 17.344 \\

3C 454.3     & -28298.74 & -28289.95 & 60.2695 & 34.617 &
\textbf{-28992.34} & \textbf{-28983.56} & \textbf{7.7698} & \textbf{8.347} \\

B2 1633+382  & -18914.01 & -18906.20 & 21.4617 & 8.945 &
\textbf{-19145.81} & \textbf{-19137.99} & \textbf{6.9984} & \textbf{3.485} \\

CTA 102      & -13694.49 & -13686.99 & 72.9160 & 68.786 &
\textbf{-14822.87} & \textbf{-14815.36} & \textbf{15.9721} & \textbf{5.619} \\

OJ 248       & -12235.51 & -12228.59 & 33.3678 & 19.695 &
\textbf{-12428.18} & \textbf{-12421.26} & \textbf{20.0341} & \textbf{13.318} \\

PKS 0736+01  & -6614.20 & -6608.42 & 12.0873 & 6.521 &
\textbf{-6751.81} & \textbf{-6746.03} & \textbf{2.8707} & \textbf{1.678} \\

PKS 1510$-$08 & -19391.62 & -19383.68 & 45.2155 & 24.272 &
\textbf{-19820.66} & \textbf{-19812.72} & \textbf{13.3601} & \textbf{4.751} \\

\hline
\end{tabular}
\end{table*}

\begin{table*}[ht]
\centering
\caption{Comparison of Gaussian, log-normal and Rice fits to the PD distribution of the blazar sample. The AIC, BIC, AD test, $\chi_r^2$, and corresponding DoF are reported for each model. The lower values of AIC, BIC, AD test, and $\chi_r^2$ corresponding to the preferred fit are highlighted in bold.}
\label{tab:4}
\setlength{\tabcolsep}{0.9pt}
\begin{tabular}{l|ccccc|ccccc|ccccc}
\hline
{Blazar name} &
\multicolumn{5}{c}{Gaussian} &
\multicolumn{5}{c}{Log-normal} &
\multicolumn{5}{c}{Rice} \\
\cline{2-6}\cline{7-11}\cline{12-16}
& AIC & BIC & AD & $\chi_r^2$ & DoF
& AIC & BIC & AD & $\chi_r^2$ & DoF
& AIC & BIC & AD & $\chi_r^2$ & DoF \\

(1) & (2) & (3) & (4) & (5) & (6) & (7) & (8) & (9) & (10) & (11) & (12) & (13) & (14) & (15) & (16) \\
\hline

Mrk 421 &
3773.79 & 3783.30 & 25.52 & 12.50 & 11 &
\textbf{3505.11} & \textbf{3514.62} & \textbf{3.60} & \textbf{2.87} & 13 & 3570.18 & 3579.69 & 24.99 & 6.99 & 11\\

PKS 2155$-$304 &
2145.63 & 2153.73 & 5.6378 & 3.29 & 10 &
2120.18 & 2128.28 & 4.0452 & 2.97 & 11 & \textbf{2075.16} & \textbf{2083.27} & \textbf{1.38} & \textbf{1.40} & 10\\

AO 0235+164 &
1898.22 & 1905.47 & 6.1031 & 5.05 & 9 &
\textbf{1831.76} & \textbf{1839.01} & \textbf{2.4106} & \textbf{2.85} & 8 & 1855.89 & 1863.14 & 12.99 & 6.37 & 9 \\

3C 273 &
-245.69 & -237.73 & 29.4189 & 32.47 & 5 &
\textbf{-591.35} & \textbf{-583.39} & \textbf{3.4848} & \textbf{7.99} & 4 & -404.27 & -396.30 & 25.89 & 21.89 & 5\\

3C 454.3 &
4430.10 & 4439.38 & 22.2515 & 19.89 & 10 &
\textbf{4165.19} & \textbf{4174.48} & \textbf{8.2028} & \textbf{4.10} & 12 & 4413.21 & 4422.49 & 91.45 & 50.44 & 9\\

B2 1633+382 &
2666.92 & 2675.10 & 18.2967 & 12.54 & 9 &
\textbf{2445.93} & \textbf{2454.11} & \textbf{0.5633} & \textbf{1.54} & 10 & 2571.11 & 2579.29 & 34.14 & 12.09 & 9\\

CTA 102 &
2757.25 & 2765.28 & 11.5872 & 12.30 & 9 &
\textbf{2607.26} & \textbf{2615.29} & \textbf{4.7595} & \textbf{2.90} & 9 & 2776.34 & 2784.37 & 58.65 & 34.24 & 9 \\

OJ 248 &
1239.98 & 1247.15 & 45.9532 & 111.78 & 4 &
\textbf{602.57} & \textbf{609.74} & \textbf{2.1469} & \textbf{6.48} & 2 & 1478.69 & 1485.86 & 299.42 & 343.51 & 3\\

PKS 0736+01 &
982.39 & 988.69 & 7.1242 & 7.71 & 6 &
\textbf{885.76} & \textbf{892.06} & \textbf{0.9438} & \textbf{1.23} & 4 & 952.76 & 959.06 & 17.35 & 10.44 & 6 \\

PKS 1510$-$08 &
2555.03 & 2563.34 & 37.5523 & 38.68 & 6 & 
\textbf{2039.76} & \textbf{2048.07} & \textbf{0.6731} & \textbf{1.48} & 6 & 2497.55 & 2505.86 & 117.57 & 68.44 & 5\\

\hline
\end{tabular}
\end{table*}

\begin{table*}[ht]
\centering
\caption{Comparison of the folded Gaussian, log-normal, and Rice distribution fits to the absolute values of the normalized Stokes parameter $|q|$ for the blazar sample. The AIC, BIC, AD test, $\chi_r^2$, and corresponding DoF are reported for each model. Lower values of AIC, BIC, AD, and $\chi_r^2$ indicate a better fit.}
% \caption{Comparison of folded Gaussian, log-normal and Rice fits to the Stokes parameter |q| distribution of the blazar sample.}
\label{tab:5}
\setlength{\tabcolsep}{0.9pt}
\begin{tabular}{l|ccccc|ccccc|ccccc}
\hline
{Blazar name} &
\multicolumn{5}{c}{Folded Gaussian} &
\multicolumn{5}{c}{Log-normal} &
\multicolumn{5}{c}{Rice} \\
\cline{2-6}\cline{7-11}\cline{12-16}
& AIC & BIC & AD & $\chi_r^2$ & DoF
& AIC & BIC & AD & $\chi_r^2$ & DoF
& AIC & BIC & AD & $\chi_r^2$ & DoF \\

(1) & (2) & (3) & (4) & (5) & (6) & (7) & (8) & (9) & (10) & (11) & (12) & (13) & (14) & (15) & (16) \\

\hline

Mrk 421 & -4946.24 & -4936.73 & 6.22 & 3.66 & 10 & -4839.67 & -4830.16 & 12.23 & 5.28 & 12 & -4462.26 & -4452.75 & 123.36 & 64.13 & 8\\

    PKS 2155$-$304 & -2019.87 & -2011.77 & 0.53 & 0.49 & 11 & -1915.09 & -1906.98 & 7.61 & 4.42 & 11 & -1926.40 & 1918.30 & 21.95 & 12.96 & 10 \\

    AO 0235+164 & -897.71 & -890.47 & 17.17 & 6.41 & 9 &
    -907.76 & -900.51 & 2.69 & 3.31 & 6 & -570.48 & -563.23 & 105.37 & 75.06 & 9\\

    3C 273 & -4555.64 & -4547.68 & 16.30 & 9.53 & 3 &
    -4602.57 & -4594.61 & 8.04 & 11.36 & 4 & -4175.15 & -4167.18 & 103.63 & 92.05 & 2 \\

    3C 454.3 & -3488.08 & -3478.79 & 32.74 & 14.10 & 8 &
    -3460.34 & -3451.05 & 8.16 & 4.18 & 10 & -2694.93 & -2685.65 & 241.35 & 177.89 & 6 \\

    B2 1633+382 & -1803.52 & -1795.34 & 18.64 & 6.90 & 8 &
    -1778.07 & -1769.89 & 6.43 & 2.77 & 8 & -1373.19 & -1365.02 & 129.01 & 74.87 & 7 \\

    CTA 102 & -1499.18 & -1491.15 & 27.10 & 10.72 & 7 & 
    -1545.65 & -1537.62 & 3.26 & 3.16 & 7 & -1014.28 & -1006.25 & 159.11 & 115.86 & 6 \\

    OJ 248 & -1704.69 & -1697.52 & 131.17 & 82.84 & 3 &
    -2100.17 & -2093.0 & 0.97 & 1.85 & 1 &-992.47 & -985.30 & 370.53 & 364.85 & 2 \\

    PKS 0736+01 & -769.53 & -763.23 & 5.78 & 2.76 & 7 & 
    -744.36 & -738.07 & 2.58 & 3.38 & 4 & -612.75 & -606.46 & 45.96 & 28.57 & 7 \\

    PKS 1510$-$08 & -2490.73 & -2482.43 & 61.50 & 29.72 & 5 & -2694.73 & -2686.62 & 2.91 & 1.79 & 5 & -1817.72 & -1809.41 & 256.50 & 187.75 & 4 \\

\hline
\end{tabular}
\end{table*}

\begin{table*}[ht]
\centering
\caption{Comparison of the folded Gaussian, log-normal, and Rice distribution fits to the absolute values of the normalized Stokes parameter $|u|$ for the blazar sample. The AIC, BIC, AD test, $\chi_r^2$, and corresponding DoF are reported for each model. Lower values of AIC, BIC, AD, and $\chi_r^2$ indicate a better fit.}
\label{tab:6}
\setlength{\tabcolsep}{0.9pt}
\begin{tabular}{l|ccccc|ccccc|ccccc}
\hline
{Blazar name} &
\multicolumn{5}{c}{Folded Gaussian} &
\multicolumn{5}{c}{Log-normal} &
\multicolumn{5}{c}{Rice} \\
\cline{2-6}\cline{7-11}\cline{12-16}
& AIC & BIC & AD & $\chi_r^2$ & DoF
& AIC & BIC & AD & $\chi_r^2$ & DoF
& AIC & BIC & AD & $\chi_r^2$ & DoF \\

(1) & (2) & (3) & (4) & (5) & (6) & (7) & (8) & (9) & (10) & (11) & (12) & (13) & (14) & (15) & (16) \\

\hline

Mrk 421 & -4879.50 & -4869.99 & 9.45 & 4.09 & 10 &
    -4716.82 & -4707.31 & 13.36 & 7.69 & 13 & -4256.09 & -4246.58 & 159.62 & 103.29 & 8 \\

    PKS 2155$-$304 & -2058.09 & -2049.99 & 5.38 & 2.16 & 9 & -1993.78 & -1985.68 & 7.79 & 4.04 & 8 & -1758.32 & -1750.22 & 75.83 & 40.52 & 7 \\

    AO 0235+164 & -944.05 & -936.79 & 1.10 & 0.91 & 8 &
    -902.17 & -894.92 & 4.07 & 3.07 & 7 & -801.52 & -794.27 & 36.28 & 21.59 & 8 \\

    3C 273 & -4142.10 & -4134.14 & 16.27 & 20.44 & 5 &
    -4103.96 & -4096.99 & 14.51 & 11.16 & 6 & -4075.36 & -4067.39 & 27.33 & 17.53 & 4 \\

    3C 454.3 & -3507.82 & -3498.53 & 28.46 & 9.93 & 10 &
    -3467.32 & -3458.04 & 8.83 & 4.89 & 12 & -2737.48 & -2728.19 & 227.12 & 145.97 & 9 \\

    B2 1633+382 & -1925.37 & -1917.19 & 6.35 & 3.01 & 10 &
    -1847.07 & -1838.89 & 8.18 & 4.45 & 9 & -1676.65 & -1668.48 & 17.24 & 15.74 & 9 \\

    CTA 102 & -1505.81 & -1497.78 & 24.19 & 9.15 & 9 &
    -1474.21 & -1466.17 & 7.47 & 5.15 & 8 & -989.71 & -981.68 & 152.62 & 112.92 & 8 \\

    OJ 248 & -1676.79 & -1669.62 & 119.04 & 93.14 & 3 &
    -2030.44 & -2023.28 & 1.74 & 6.19 & 1 & -1022.50 & -1015.33 & 343.28 & 310.81 & 3 \\

    PKS 0736+01 & -836.29 & -829.99 & 8.67 & 6.09 & 4 &
    -860.01 & -853.72 & 2.32 & 6.62 & 3 & -654.77 & -648.47 & 55.64 & 53.14 & 4 \\

    PKS 1510$-$08 & -2393.91 & -2385.59 & 43.09 & 21.38 & 5 & -2539.29 & -2530.98 & 3.38 & 2.94 & 5 & -1817.61 & -1809.29 & 205.55 & 143.96 & 4 \\
\hline
\end{tabular}
\end{table*}

\subsubsection{Rms-flux relation}

\begin{figure}[ht!]
\centering
\includegraphics[width=0.45\textwidth]{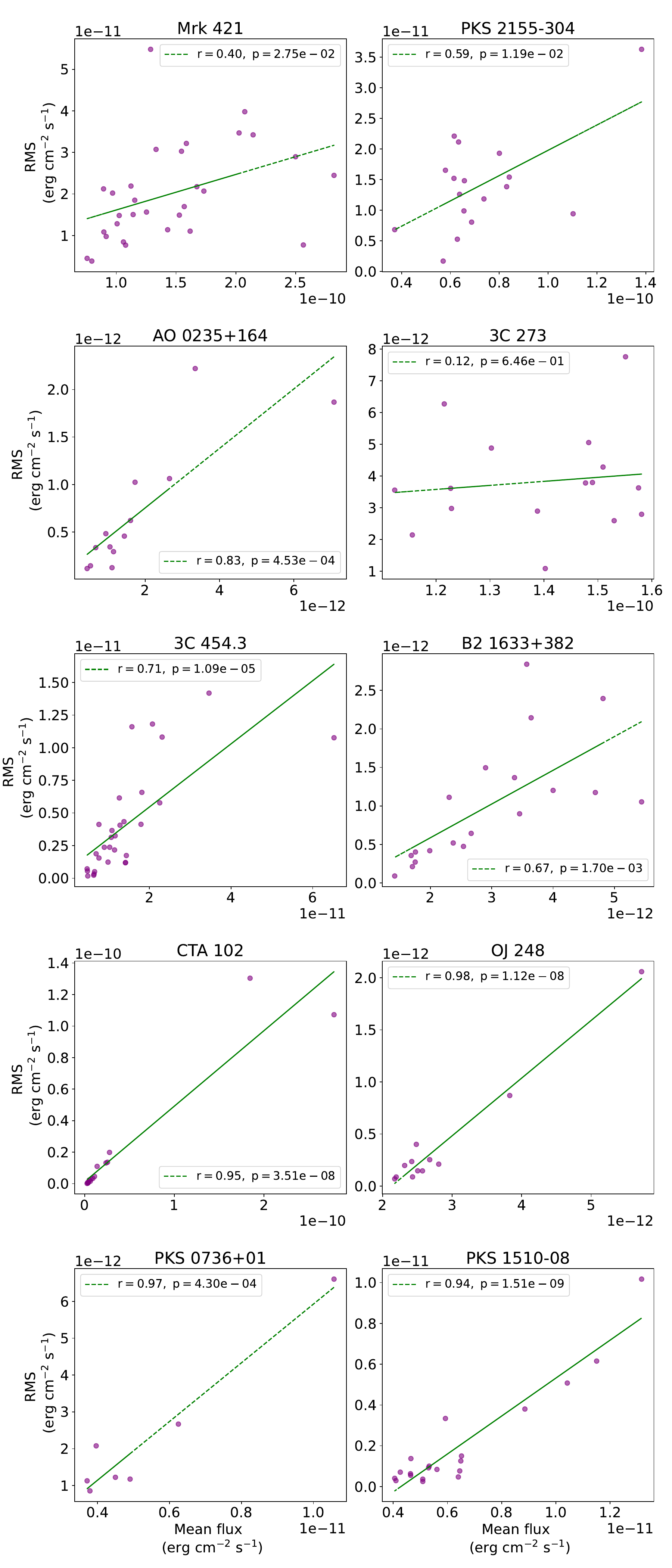}
\caption{The optical rms–ﬂux diagram for the sample blazars.}
\label{fig:5}
\end{figure}

\begin{figure}[ht!]
\centering
\includegraphics[width=0.45\textwidth]{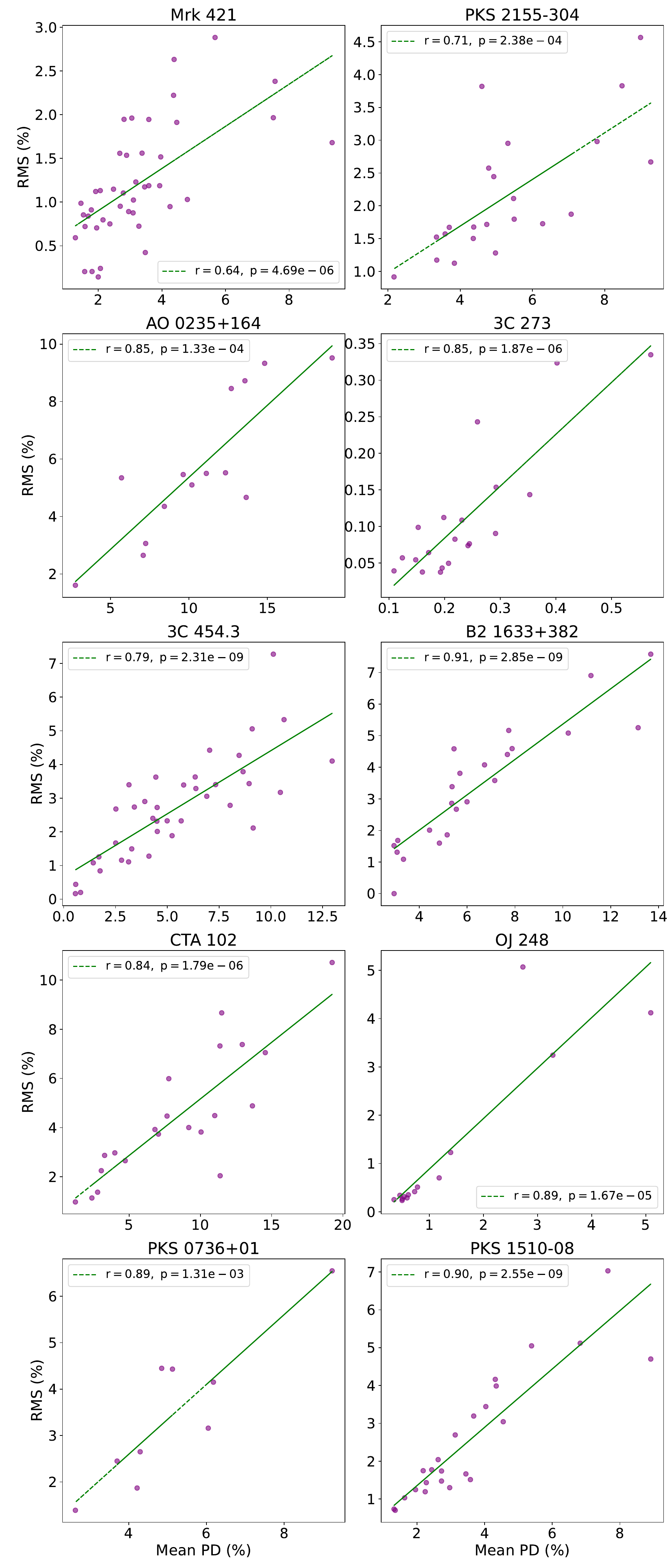}
\caption{The optical rms–PD diagram for the sample blazars.}
\label{fig:6}
\end{figure}

To further investigate the variability characteristics of the sample of blazars, we examined the correlation between the mean emission level and the amplitude of variability by plotting the rms-flux and rms-PD relations. These relations quantify how the intrinsic variability scales with the average state of the source and serve as important diagnostics of the underlying stochastic processes that govern the emission. For this purpose, the observed light curves were divided into N segments of equal length. For statistical reliability, each segment should contain at least 20 data points. For every segment, we calculated the mean optical flux (mean PD) along with the corresponding root mean square (rms) variability \citep{Wang_2023}. The resulting rms-flux and rms-PD graphs are presented in Figures \ref{fig:5} and \ref{fig:6}. We also calculated the Pearson correlation coefficient (r) between the emission level and the variability amplitude for each plot. A positive correlation between rms and flux, widely observed in accreting systems, is generally interpreted as evidence for multiplicative variability processes rather than additive ones. This naturally gives rise to a log-normal distribution. Extending this framework to the PD, a similar linear relation between the rms and PD may imply that turbulence in the magnetic-field and particle acceleration mechanisms are coupled in a multiplicative manner, thereby imprinting similar statistical behavior on both flux and polarization variability.

\subsubsection{Behavior of normalized Stokes parameters}

\begin{figure}[ht!]
\centering
\includegraphics[width=0.47\textwidth]{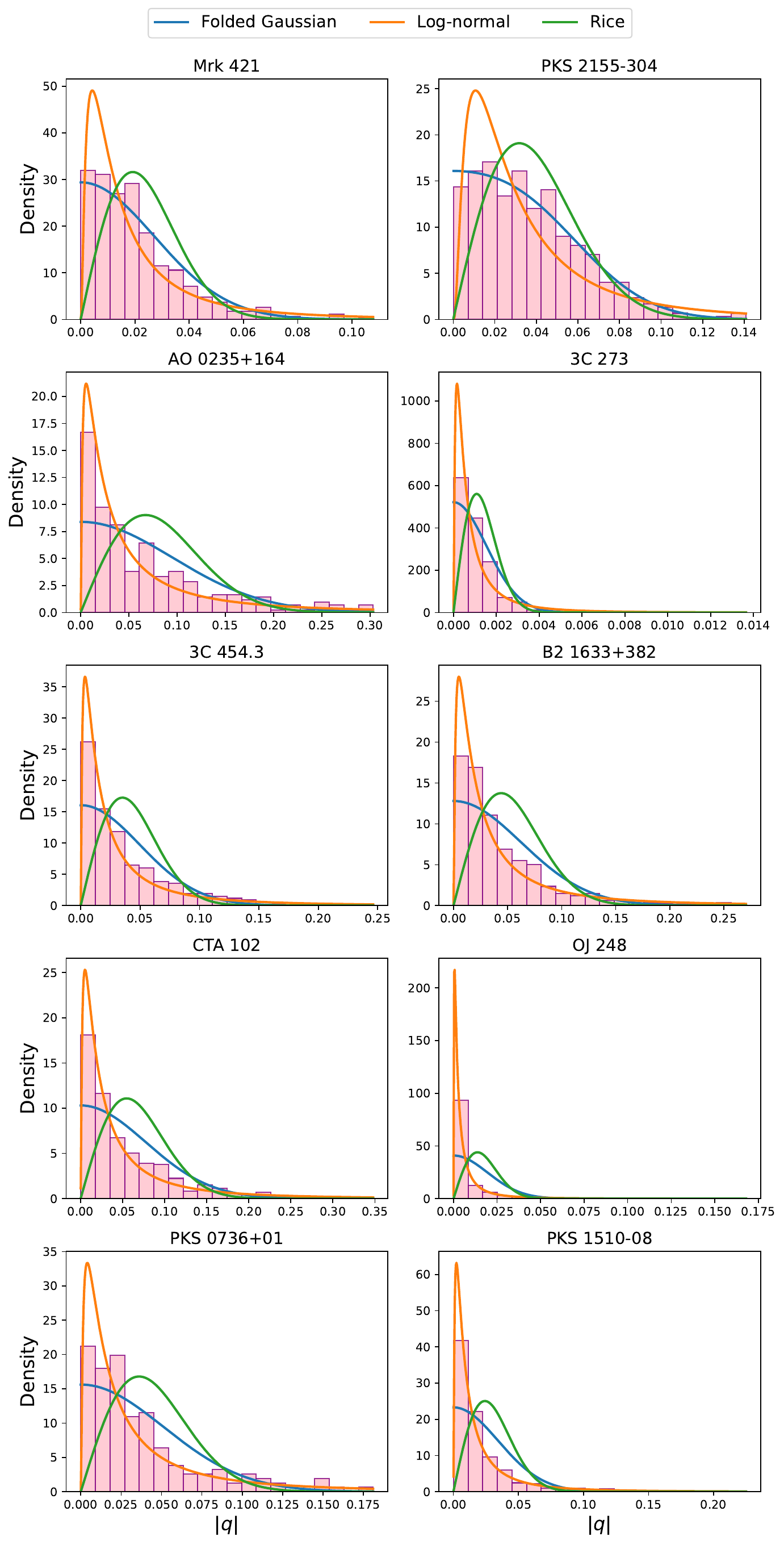}
\caption{Stokes parameter |q| distribution of the sample of blazars. The blue, orange and green lines correspond to folded Gaussian, log-normal and Rice distribution, respectively.}
\label{fig:7}
\end{figure}

\begin{figure}[ht!]
\centering
\includegraphics[width=0.45\textwidth]{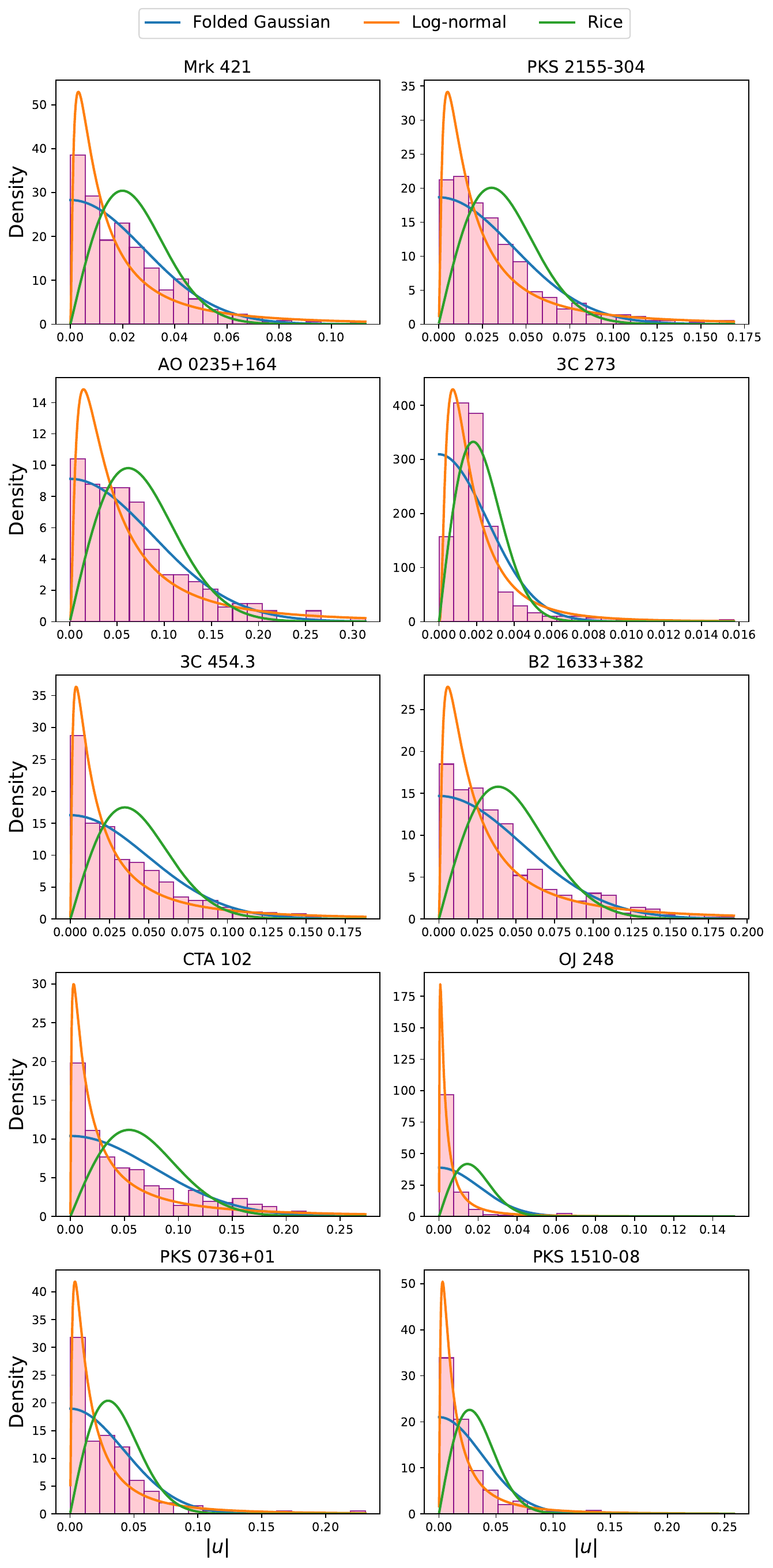}
\caption{Stokes parameter |u| distribution of the sample of blazars. The blue, orange and green lines correspond to folded Gaussian, log-normal and Rice distribution, respectively.}
\label{fig:8}
\end{figure}

We also fit the distribution of normalized Stokes parameters, $|q|$ and $|u|$, to investigate whether their variability is consistent with a log-normal distribution. Since the log-normal distribution is defined only for positive-valued quantities, we perform the analysis using the absolute values, $|q|$ and $|u|$, as the estimated values of PD remain unchanged. The resulting distributions are shown in Figures \ref{fig:7} and \ref{fig:8}. In addition, we also fit the folded-Gaussian and Rice distribution to the Stokes-parameter distributions. The estimated results are reported in Tables \ref{tab:5} and \ref{tab:6}, respectively.

\section{Discussion}
\label{sec:6}

From the above mentioned statistical analysis, we find that in nearly all cases the distributions of optical flux and PD (except 3C 273 and PKS 2155-304) are better described by log-normality over normality. Such behavior suggests that the observed variability is dominated by a sequence of many non-linear multiplicative fluctuations operating within the relativistic jet rather than by the linear superposition of these independent emission regions \citep{Uttley_2005}. For optical synchrotron flux distribution, several studies suggest that this interpretation is relatively straightforward. The observed emission depends on many independent physical scenarios, including the number density of relativistic electrons, the magnetic-field fluctuations, the particle acceleration efficiency, the Doppler factor, and the emitting volume \citep{Rybicki_1985, :Longair_2011, Urry_1995}. 
The physical origin of the log-normal distribution of PD is less obvious because polarization is a vector quantity rather than a scalar intensity. Nevertheless, PD is determined by the ordering of the magnetic-field within the synchrotron emitting region \citep{Lyutikov_2005}. So, any physical scenario that can cause fluctuations in the magnetic-field also causes fluctuations in PD. Unlike additive random walks of polarization vectors, the magnetic-field inside relativistic jets evolves through coupled non-linear magneto-hydrodynamic (MHD) processes, implying that successive changes in polarization are not statistically independent but instead modulate over one another. In non-ideal MHD, the magnetic-field evolves as \citep{Priest_2014, Kulsrud_2004, Goedbloed_2019},

\begin{equation}\label{eq:8}
\rm    \frac{\partial \vec{B}}{\partial t} = \vec{\nabla} \times (\vec{v} \times \vec{B}) + \eta \vec{\nabla}^2 \vec{B}
\end{equation}

where the parameter $\eta$ is the magnetic diffusion coefficient and $\vec{v}$ is the plasma velocity.

The first term on the right-hand side of the Eq \ref{eq:8} describes the fluctuations in the magnetic-field structure by the plasma flow through advection, twisting, and compression, whereas the second term represents magnetic diffusion and dissipation effects caused by finite electrical resistivity. Now, since relativistic jets are characterized by very large magnetic Reynolds number, diffusion is generally much weaker than advection, and hence magnetic-field remains effectively frozen into the plasma \citep{Priest_2014, Kulsrud_2004}. As a result, the field is continuously reshaped by the bulk motion of the jet. As the plasma blob propagates downstream, the magnetic-field is repeatedly modified by shocks, turbulent motions, and velocity shear \citep{Marscher_2014, Hughes_1985, Hughes_1989a, Lyutikov_2005}. Importantly, each new disturbance acts on a magnetic configuration that has already been altered by previous events rather than on the original field. Consequently, the field evolution is better described as a sequence of successive fractional changes than as a simple sum of independent perturbations. This can be written schematically as,

\begin{equation}
\rm    \vec{B}_{n+1} = \vec{B}_n \ (1 + \delta_n)
\end{equation}

where $\delta_n$ denotes the fractional change produced during $n^{th}$ interaction. After many such events, the magnetic-field strength becomes

\begin{equation}
 \rm   \vec{B} = \vec{B}_0 \ \prod_{i=1}^{N} (1 + \delta_i)
\end{equation}

showing that the cumulative evolution of the field is inherently multiplicative.

Taking the logarithm of above equation gives,

\begin{equation}
 \rm   ln\ \vec{B} = ln\ \vec{B}_0 \ + \sum_{i=1}^{N} ln\ (1+\delta_i)
\end{equation}

If the fractional perturbations are statistically independent, the CLT predicts the sum of the right hand side of a fore-mentioned equation follows a Gaussian distribution. As a result, the magnetic-field strength itself is expected to follow a log-normal distribution. Such behavior naturally arises from the cascading effect of several small stochastic disturbances and does not require any individual event to dominate the evolution.

Since optical synchrotron polarization depends on both the strength and the ordering of the magnetic-field, fluctuations in the magnetic-field are expected to be reflected in the observed PD. In a transverse shock scenario, shocks compress and partially align the magnetic-field, increasing both the synchrotron emissivity and PD \citep{Marscher_Gear_1985, Hughes_1985}. In helical magnetic-field model, plasma moving through an ordered helical field changes the projected field geometry, producing systematic variations in polarization \citep{Lyutikov_2005, Marscher_2008, Blandford_1977, Blandford_1983}. Similarly, in the TEMZ model, turbulent cells with different magnetic-field orientations repeatedly cross standing shocks, continuously modifying the magnetic structure inherited from earlier interactions \citep{Marscher_2014}. Although these models differ in their physical details, they all involve repeated evolution of the magnetic-field through stretching, compression, or re-orientation. The MHD induction equation therefore provides a common physical framework in which these successive perturbations naturally generate multiplicative variability. This offers a plausible explanation for why both the optical flux and, more importantly, PD of most blazars in our sample are better described by log-normal rather than Gaussian distributions.

Many authors suggest that the presence of log-normal distributions in astrophysical sources is often associated with the multiplicative variability processes operating within the system. However, studies by \citet{Giebels_2012} suggest that log-normal behavior may also arise from certain classes of additive stochastic processes, implying that the presence of log-normality does not uniquely imply multiplicative mechanisms.

\section{Conclusion}
\label{sec:7}

In this work, we have carried out an in-depth temporal analysis of the decade-long optical emission from a selected sample of blazars. By combining variability analysis, probability distribution, and rms-flux (rms-polarization) relations, we have explored the statistical nature of the variability and its possible physical origin. The optical emission of all sources is found to be intrinsically variable over long timescales. More importantly, both the optical flux and PD are generally better described by log-normal rather than other fitted distribution. The agreement among the AIC, BIC, AD, and $\rm \chi_r^2$ statistics, together with the presence of linear rms–flux and rms–polarization relations, suggest that the observed variability is more naturally explained by multiplicative stochastic processes than by the simple addition of independent fluctuations. 

A notable result of this study is that the PD shows statistical properties that are remarkably similar to those of the synchrotron flux. Because polarization is directly related to the magnetic-field structure in the emission region, this similarity suggests that the long-term evolution of the magnetic-field may also be driven by cumulative stochastic processes. Within the MHD framework, repeated stretching, compression, and re-orientation of the magnetic-field by shocks, turbulence, velocity shear, and large-scale jet motions provide a plausible mechanism for producing multiplicative variability. These processes can affect both the synchrotron intensity and PD, indicating that the two observables may share a common physical origin over long timescales. However, the presence of log-normal distributions alone does not uniquely determine the underlying mechanism, since similar statistical behavior can also emerge in some additive stochastic models. Distinguishing between these possibilities will require a larger blazar sample, multi-wavelength polarization monitoring, and detailed numerical MHD simulations. Overall, our results show that optical polarization contains statistical signatures that are closely linked to the evolution of the jet magnetic-field. The frequent occurrence of log-normal behavior in both optical flux and polarization provides further observational support for the role of multiplicative MHD processes in shaping the long-term variability of relativistic jets. It also establishes optical polarization as a useful statistical probe of magnetic-field evolution and particle acceleration in blazars.

\section*{Acknowledgements}

Authors thank the anonymous reviewer for his/her critical comments and important suggestions to improve the contents of the manuscript scientifically. Data from the Steward Observatory spectro-polarimetric monitoring project were used. This program is supported by Fermi Guest Investigator grants NNX08AW56G, NNX09AU10G, NNX12AO93G, and NNX15AU81G. ACG thankfully acknowledges support from the Board of Research in Nuclear Sciences (BRNS), Department of Atomic Energy (DAE) funded project (Project No. 57/14/06/2025-BRNS/727 Dated 16-09-2025). A. Singh also thanks Mr. Atul Pathania for scientific discussions and suggestions.

% \appendix

% \section{Appendix title 1}
% %% \label{}

% \section{Appendix title 2}
% %% \label{}

\bibliographystyle{elsarticle-harv} 
\bibliography{main}



%% For example:

%% \bibitem[Aladro et al.(2015)]{Aladro15} Aladro, R., Martín, S., Riquelme, D., et al. 2015, \aas, 579, A101

%%\end{thebibliography}

\end{document}